\documentclass[aps,twocolumn]{revtex4}
\usepackage[utf8]{inputenc}
 \usepackage{amsmath,amssymb}
\usepackage{color}                                           
\begin{document}
\title{Emergence of the cosmic space inspired by mass-to-horizon entropy}
\author{Ahmad Sheykhi\footnote{asheykhi@shirazu.ac.ir}}
\address{Department of Physics, College of
Science, Shiraz University, Shiraz 71454, Iran\\
Biruni Observatory, College of Science, Shiraz University, Shiraz
71454, Iran}

\begin{abstract}

The conception of gravity as an emergent phenomenon, rooted in the
thermodynamics of spacetime, offers a radical departure from its
geometric description. This paper investigates the emergence of
cosmic space by synthesizing two key thermodynamic approaches: the
equilibrium perspective, where the first law of thermodynamics is
applied to the apparent horizon, and the dynamic perspective of
Padmanabhan, where the cosmic space emerges as cosmic time
progresses. The central element of our study is the incorporation
of a mass-to-horizon entropy relation, $M=\gamma {c^2 L^n}/{G}$,
where $M$ denotes the effective mass associated with the system,
$L$ corresponds to the cosmological horizon, and $\gamma$ is a
constant with dimensions $[L]^{1-n}$. We first use this relation
within the Clausius relation and apply the first law of
thermodynamics, $dE=T_h dS_h+WdV$, on the apparent horizon to
derive the modified Friedmann equations. Subsequently, we embed
the mass-to-horizon entropy relation into Padmanabhan's cosmic
emergence proposal, the dependence of the volume change on the
degrees of freedom in the bulk and on the boundary, and show its
consistency with the thermodynamically derived equations. The
successful reconstruction of the modified Friedmann equations
through these independent yet convergent thermodynamic routes
strongly suggests that the mass-to-horizon entropy is a
fundamental bridge between the information-theoretic
microstructure of spacetime and its effective cosmological
description. Finally, we show that the generalized second law of
thermodynamics is fulfilled for the universe enveloped by the
apparent horizon.

\end{abstract}
 \maketitle

 \newpage
\section{Introduction\label{Intro}}
The conceptual foundation of General Relativity (GR), which
interprets gravity as the curvature of a classical spacetime
manifold, has been supremely successful on astrophysical and
cosmological scales. However, its inherent clash with the
principles of quantum mechanics in regimes such as the primordial
universe and black hole singularities necessitates a more profound
underlying theory. Among the most intriguing clues guiding this
search is the remarkable connection between the laws of gravity
and the laws of thermodynamics. This connection was first
crystallized in the context of black hole mechanics \cite{Haw1},
where Bardeen, Carter, and Hawking established analogs of the four
laws of thermodynamics, with the horizon area playing the role of
entropy and surface gravity that of temperature. Hawking's seminal
work \cite{Haw2} later cemented this analogy by demonstrating that
black holes indeed radiate with a temperature proportional to
surface gravity, $T_{BH} = {\kappa}/{2\pi}$, solidifying the
Bekenstein-Hawking entropy $S_{BH}= A/4 G$. Throughout this work
we choose the units as $\hbar=c=k_B=1$.

This thermodynamic-gravity connection extends far beyond
stationary black holes. Jacobson's groundbreaking work \cite{Jac}
demonstrated that the Einstein field equation itself can be
derived from the Clausius relation, $\delta Q=TdS$, applied to
local Rindler horizons, assuming the entropy is proportional to
the horizon area. This result provided a compelling argument that
gravity is not a fundamental force and can be understood through
thermodynamic arguments \cite{Pad1,Pad2,Pad3}. In the cosmological
context, this paradigm was powerfully applied by assuming the
first law of thermodynamics holds on the apparent horizon of a
Friedmann-Robertson-Walker (FRW) universe, leading to the
successful derivation of the Friedmann equations
\cite{CaiKim,Cai2,Shey1,Shey2}. The apparent horizon, a causal
boundary defined by the condition, $h^{\mu
\nu}\partial_{\mu}\tilde{r}_A\partial_{\nu}\tilde{r}_A=0$ (where
$\tilde{r}_A=a(t) r$), is a suitable boundary from thermodynamic
perspective.

A distinct, yet deeply related, perspective on the emergence of
spacetime was proposed by Padmanabhan \cite{PadEm}. He argued that
the expansion of the cosmos, encoded in the evolution of the
cosmic horizon, can be understood as the process of its
microscopic degrees of freedom coming into equilibrium with those
in the bulk. His "emergent gravity" paradigm is encapsulated in
the dynamical equation
\begin{equation}
\frac{dV}{dt}\propto
\left(N_{\mathrm{sur}}-N_{\mathrm{bulk}}\right), \label{dV1}
\end{equation}
where $V$ is the volume of space, $N_{\mathrm{sur}}$  is the
number of surface degrees of freedom on the horizon and
$N_{\mathrm{bulk}}$ is the number of degrees of freedom related to
the Komar energy in the enclosed volume. This approach not only
reproduces the standard Friedmann equations but also provides a
compelling narrative for the emergence of space from a
pre-geometric state \cite{CaiEm,Yang1,Sheyem}.

A critical implication of this thermodynamic/emergent gravity
framework is that any modification to the Bekenstein-Hawking
entropy-area law, as expected from quantum gravity (e.g., string
theory, loop quantum gravity), must inevitably lead to
modifications of the gravitational field equations
\cite{Cai3,sheyECFE}. Entropy corrections such as the logarithmic
\cite{SheyLog} or power-law \cite{SheyPL} terms have been
extensively studied, leading to modified Friedmann equations with
extra terms that can mimic dark energy or influence early universe
inflation.

A cosmology-centric critique \cite{Goh1,Goh2} reveals a
significant constraint in constructing entropic models of gravity.
The argument shows that when two conditions are met: (i) the
Clausius relation defines the horizon temperature to ensure
thermodynamic consistency, and (ii) the mass-horizon relation
(MHR) is linear-the resulting cosmological model becomes
indistinguishable from the standard one based on
Bekenstein-Hawking entropy. This forces all such models to share
the same shortcomings, including an inability to accurately match
the observed expansion history and growth of cosmic structures
\cite{Bas1,Bas2}. To overcome this fundamental constraint, a
generalized mass-horizon relation has been proposed, which
naturally leads to a modified entropy that encompasses forms like
Tsallis-Cirto and Barrow entropy as specific cases \cite{Tsa,Bar}.

Recent work \cite{Goh2} has shown that the generalized
mass-horizon entropy framework can yield a cosmological model
that, for certain parameter values, fits observational data as
well as the standard $\Lambda$CDM model. Furthermore, by applying
the gravity-thermodynamics conjecture, the modified Friedmann
equations derived from this entropy \cite{Bas3} naturally
incorporate an effective dark energy component, which originates
from the extra terms in the generalized entropy expression. The
implications of the modified cosmology inspired by mass-to-horizon
entropy to the growth of matter perturbations within the spherical
Top-Hat formalism in the linear regime, and primordial
gravitational waves has been explored recently in \cite{Luci1}.
Very recently, the authors of \cite{Luci2} observationally
constrained the modified mass-to-horizon cosmological model using
a combination of Type Ia supernovae (SNIa), cosmic chronometers
(CC), and baryon acoustic oscillations (BAO) data, including the
Second Data Release of the Dark Energy Spectroscopic Instrument
(DESI DR2) survey, together with the Supernovae $H_0$ for the
Equation of State (SH0ES) distance-ladder prior, across four
combinations of data sets. They argued  that the best-fit value
for the entropic exponent $n$ is found to be less than unity,
whereas the corresponding estimate for $\gamma$ exceeds unity
\cite{Luci2}.

In the present work, we construct the modified Friedmann equations
inspired by a general mass-to-horizon entropy relation. We first
show that starting from the first law of thermodynamics, one is
able to translate it to the first Friedmann equation on the
apparent horizon. We then embed this entropy relation into
Padmanabhan's emergence scenario. By re-formulating the surface
degrees of freedom in terms of the mass-horizon entropy, we will
derive the cosmic evolution from the principle of emergence. The
consistency of the results obtained from these two independent
approaches-the equilibrium thermodynamics of the horizon and the
dynamic process of emergence-will provide a robust and
cross-validated framework.

This paper is structured as follows. In Section \ref{MH}, we
address the question why we should consider the generalized
mass-to horizon relation? In Section \ref{First} we explore
thermodynamic setup of the FRW universe on the apparent horizon,
constructing modified Friedmann equations through first law of
thermodynamics. In Section \ref{Emer}, we reconstruct the same
equations from the perspective of Padmanabhan's emergence
proposal, using the same entropy ansatz. Section \ref{cosm} is
devoted to a discussion of the physical implications and
cosmological consequences of our derived modifications. Finally,
we present our conclusions in Section \ref{Con}.

\section{Why generalized mass-to-horizon entropy? \label{MH}}
In this section, we review the main motivations for considering
the generalized mass-to-horizon entropy in the context of
thermodynamics-gravity conjecture, assuming non-extensive entropy.
We follow the arguments given in \cite{Goh2}. The application of
thermodynamics to the cosmos is founded on the holographic
principle \cite{Hooft,Sus}. This principle, a generalization of
black hole thermodynamics, asserts that for a universe with a
cosmological horizon, the information within the bulk volume can
be represented by degrees of freedom on its two-dimensional
boundary. This framework allows us to assign standard
thermodynamic quantities to the horizon itself. A consistent
formulation requires three key elements:
\begin{itemize}
\item Holographic Association: The entropy ($S$), mass ($M$), and energy
($E$) must be properties of the cosmic horizon.
\item Thermodynamic Law: These quantities are linked by the Clausius
relation: $dE =c^2 dM =TdS$, where $T$ is the Hawking temperature.
\item Geometric Relation: A linear mass-to-horizon relation (MHR) is
assumed: $M=\frac{c^2}{G} L$, where $L$ is the cosmological
horizon.
\end{itemize}
The term "consistent" here denotes a set of thermodynamic
assumptions that collectively reproduce the standard
Bekenstein-Hawking entropy. These are the identifications $E = M$
and $T =T_{H}$, a linear mass-horizon relation, and adherence to
the Clausius relation, $TdS=dE$.

This logic dictates that the relation $E = M$ should itself emerge
from the Clausius relation and the holographic definitions of $T$
and $S$. The standard framework is trivially consistent, but a
significant issue emerges with non-extensive entropies. As
established in \cite{Noj,Cimd1}, combining such entropies with the
Hawking temperature in the Clausius relation leads to an
inconsistent mass-energy relation, revealing a fundamental
thermodynamic incompatibility between these elements \cite{Cimd1}.

The application of the holographic principle naturally leads to a
key question: is it possible to utilize non-extensive entropies
instead of the Bekenstein-Hawking formula? To address this, we
consider two mutually exclusive strategies:

(i) Derive a new temperature: Adhere strictly to the Clausius
relation and a linear mass-horizon relation. For a chosen
non-extensive entropy, this framework defines a corresponding
horizon temperature that ensures consistency \cite{Cimd2}. The
drawback is that these derived temperatures lack the robust
justification of the Hawking temperature, as they are not
supported by quantum field theory in curved spacetime.

(ii) Keep Hawking temperature and redefine other relations:
Acknowledge that the Hawking temperature, grounded in surface
gravity, is the physically preferred choice. In this case, one
must abandon the linear mass-horizon relation to maintain
consistency within the Clausius relation. The objective becomes to
construct a new, consistent thermodynamic framework where
non-extensive entropy and the Hawking temperature coexist without
contradiction.

Here we propose a generalized, non-linear MHR. By applying this
relation alongside the Hawking temperature, we derive a new,
thermodynamically consistent definition of horizon entropy that
aligns with the holographic principle. We then implement this new
entropy within the framework of cosmology and construct the
modified dynamical equations describing the evolution of the
universe. In this framework, entropic force terms are added to the
Einstein field equations to explain the universe's accelerated
expansion, while GR itself is not altered. It is essential to
distinguish this from Verlinde's entropic gravity \cite{Ver}. Our
model is an extension of classical GR, whereas Verlinde's proposes
that gravity is entirely an emergent entropic phenomenon.

The primary motivation for a new MHR stems from a critical finding
in \cite{Goh1}: any entropic model that assumes a linear MHR and
enforces thermodynamic consistency via the Clausius relation will
inevitably reproduce the standard entropic force derived from
Bekenstein entropy and Hawking temperature. Consequently, such
models are fundamentally constrained to inherit the same
limitations as the standard framework, including its failures to
accurately describe both the cosmological background evolution and
the growth of perturbations \cite{Bas1,Bas2}.

Building on the result from \cite{Goh1} that the entropic force
depends critically on the form of the MHR, we take the logical
step of generalizing the MHR itself. We therefore propose and
investigate the following generalized relation \cite{Goh2}
\begin{equation}\label{MHR}
M=\gamma \frac{c^2}{G}L^n,
\end{equation}
where $n$ is a non-negative real number, and $\gamma$ is a
constant with dimensions $[L]^{1-n}$. This generalization is a
crucial prerequisite for defining thermodynamically consistent
quantities on the cosmological horizon. Its geometric validity is
supported for the specific case $n = 1$, $\gamma=1/2$, where it
reduces to the Misner-Sharp mass for the apparent horizon in
spherical symmetry \cite{Yang}. The work in \cite{Yang} further
demonstrates that such a general mass-like function is essential
for linking the geometry of the horizon (via a geometrical first
law) to the Friedmann equations, thereby validating the use of
thermodynamic concepts like the linear MHR in the standard
Bekenstein-Hawking framework.

While the case $n=1$ is geometrically justified by GR, our
generalized form ($n\neq1$) currently lacks a similar foundational
derivation. Although \cite{Yang} shows that generalized mass
functions appear in other theories of gravity, a full geometric
justification for our ansatz remains a subject for future work.
The reliability of Eq. (\ref{MHR}) has been ultimately tested
against observational data \cite{Goh2}. It was shown that
for$n=3$, the cosmological model derived from the generalized
mass-horizon entropy becomes fully equivalent to the standard
$\Lambda$CDM model. This equivalence offers a novel thermodynamic
perspective on the origin and nature of the cosmological constant.

By combining the generalized mass-horizon relation (\ref{MHR})
with the Hawking temperature in the Clausius relation, we derive a
new entropy associated with the cosmological horizon as
\cite{Goh2}
\begin{equation}\label{Sh}
S_h=\gamma \frac{2n}{n+1}L^{n-1} S_{BH},
\end{equation}
where $S_{BH}$ is the usual Bekenstein-Hawking entropy which obeys
the area law and $L$ is the cosmological radius. Crucially, this
generalized form recovers the standard framework when
$\gamma=n=1$, yielding both the linear MHR and $S_h=S_{BH}$. This
generalized entropy formula provides the necessary flexibility to
encompass several other well-known entropy proposals. For
instance: (i) Setting $n=2\delta-1$ recovers the non-extensive
Tsallis-Cirto entropy \cite{Tsa}. (ii) Setting $n=1+\Delta$, where
$0\leq\triangle\leq1$ yields Barrow entropy \cite{Bar}, implying a
compatible parameter range of $1\leq n\leq2$. (iii) For $n=d-1$ it
recovers Tsallis-Zamora entropy for cosmic horizons \cite{Zam}.
\section{Modified Friedmann equation from first law of thermodynamics\label{First}}
Our starting point is a spatially homogeneous and isotropic FRW
universe which is described by the line elements
\begin{equation}
ds^2={h}_{\mu \nu}dx^{\mu}
dx^{\nu}+\tilde{r}^2(d\theta^2+\sin^2\theta d\phi^2),
\end{equation}
where $a(t)$ is the scale factor, $\tilde{r}=a(t)r$, and $k$ is
the curvature parameter which indicates open, flat, and closed
universes, for $k = -1,0, 1$, respectively. Here we take $x^0=t,
x^1=r$, and $h_{\mu \nu}$=diag $(-1, a^2/(1-kr^2))$. In cosmology
we have several horizon, but the most well-known and consistent
from thermodynamical view point is the apparent horizon. In the
background of FRW universe, the radius of the apparent horizon is
determined via $h^{\mu
\nu}\partial_{\mu}\tilde{r}_A\partial_{\nu}\tilde{r}_A=0$, which
implies that the vector $\nabla \tilde{r}_A$ is null on the
apparent horizon surface. We find \cite{Hay1,Hay2,Bak}
\begin{equation}\label{radius}
 \tilde{r}_A=\frac{1}{\sqrt{H^2+k/a^2}},
\end{equation}
where $H=\dot{a}/a$ is the Hubble parameter. Consider the universe
as a thermodynamical system with apparent horizon as its boundary.
Similar to black hole thermodynamics, we can associate a surface
gravity and hence a temperature to the apparent horizon. The
surface gravity of the apparent horizon is defined as
\cite{Hay1,Hay2,Bak}
\begin{equation}\label{kappa}
\kappa = \frac{1}{2\sqrt{-h}} \partial_{\mu} \left( \sqrt{-h}
h^{\mu \nu}
\partial_{\nu} \tilde{r}_A \right).
\end{equation}
The associated temperature with the apparent horizon is obtained
as \cite{Hay1,Hay2,Bak}
\begin{equation}\label{T}
T_h=\frac{\kappa}{2\pi}=-\frac{1}{2 \pi \tilde
r_A}\left(1-\frac{\dot {\tilde r}_A}{2H\tilde r_A}\right).
\end{equation}
We posit that the universe's matter and energy are represented as
a perfect fluid, characterized by the energy-momentum tensor given
by:
\begin{equation}\label{TEM}
T_{\mu\nu}=(\rho+p)u_{\mu}u_{\nu}+pg_{\mu\nu},
\end{equation}
where $\rho$ denotes the energy density and $p$ represents the
pressure. The conservation of total matter and energy in the
universe is expressed by the equation, $\nabla_{\mu}T^{\mu\nu}=0$.
In the context of FRW geometry, this conservation law translates
to
\begin{equation}\label{Cons}
\dot{\rho}+3H(\rho+p)=0.
\end{equation}
Additionally, the work density associated with the universe's
volume change is defined accordingly \cite{Hay2}
\begin{equation}\label{Work}
W=-\frac{1}{2} T^{\mu\nu}h_{\mu\nu},
\end{equation}
which leads to
\begin{equation}\label{Work2}
W=\frac{1}{2}(\rho-p).
\end{equation}
To derive the Friedmann equations from the thermodynamics-gravity
conjecture, we start by invoking the first law of thermodynamics
on the apparent horizo,
\begin{equation}\label{FL}
dE = T_h dS_h + WdV.
\end{equation}
The total energy of the universe contained within the apparent
horizon is expressed as $E=\rho V$, where
$V=\frac{4\pi}{3}\tilde{r}_{A}^{3}$ represents the volume.
Additionally, $T_h$ and $S_h$ denote the temperature and entropy
associated with the apparent horizon, respectively. It can be
readily demonstrated that,
\begin{equation}\label{dE1}
 dE=4\pi\tilde
 {r}_{A}^{2}\rho d\tilde {r}_{A}+\frac{4\pi}{3}\tilde{r}_{A}^{3}\dot{\rho} dt.
\end{equation}
Using the conservation equation (\ref{Cons}), we find
\begin{equation}
\label{dE2}
 dE=4\pi\tilde
 {r}_{A}^{2}\rho d\tilde {r}_{A}-4\pi H \tilde{r}_{A}^{3}(\rho+p) dt.
\end{equation}
We propose the entropy of the apparent horizon is in the form of
the generalized mass-to-horizon entropy,
\begin{equation}\label{Ent1}
S_{h}=\frac{2\pi n \gamma}{G(n+1)}\tilde {r}_{A}^{n+1}.
\end{equation}
Taking differential form of the entropy (\ref{Ent1}), we arrive at
\begin{eqnarray} \label{dSh}
dS_h&=& \frac{2\pi n \gamma}{G}\tilde {r}_{A}^{n} d\tilde {r}_{A}.
\end{eqnarray}
Substituting relations (\ref{T}), (\ref{Work2}), (\ref{dE2}) and
(\ref{dSh}) in the first law of thermodynamics, (\ref{FL}), after
some algebra, we find the differential form of the Friedmann
equation as
\begin{equation} \label{Fried1}
n \gamma \tilde {r}_{A}^{n-4} d\tilde {r}_{A}= 4 \pi G H(\rho+p)
dt.
\end{equation}
Using the continuity equation, we arrive at
\begin{equation} \label{Fried2}
-2 n\gamma \tilde {r}_{A}^{n-4} d\tilde {r}_{A}= \frac{8 \pi G}{3}
d \rho.
\end{equation}
After integrating, we reach
\begin{equation} \label{Fried3}
\frac{2 n\gamma}{3-n} \tilde {r}_{A}^{n-3}  =  \frac{8 \pi
G}{3}\left(\rho+\rho_{\Lambda}\right),
\end{equation}
where $\Lambda$ serves as an integration constant that can be
interpreted as the cosmological constant, and
$\rho_{\Lambda}=\Lambda/(8\pi G)$. Substituting $\tilde {r}_{A}$
from Eq.(\ref{radius}), we arrive at
\begin{equation} \label{Fried4}
\left(H^2+\frac{k}{a^2}\right)^{(3-n)/2} = \frac{4\pi G}{3 n
\gamma}(3-n) (\rho+\rho_{\Lambda}).
\end{equation}
If we define an effective gravitational constant as,
\begin{equation} \label{Gef}
G_{\rm eff}=\frac{(3-n) G}{2 n \gamma},
\end{equation}
we can rewrite the modified Friedmann equation as
\begin{equation} \label{Fried5}
\left(H^2+\frac{k}{a^2}\right)^{(3-n)/2} = \frac{8\pi G_{\rm
eff}}{3} (\rho+\rho_{\Lambda}).
\end{equation}
When $n=\gamma=1$, one finds $G_{\rm eff}\rightarrow G$ and the
Friedmann equation (\ref{Fried5}) restores the result of standard
cosmology, as expected. Thus in comparison to the standard
cosmology here, we have two new parameters $n$ and $\gamma$. These
parameters can be constrained using cosmological observational
data. On the other hand for $n=\triangle+1$, the Friedman equation
in Barrow cosmology is restored \cite{SheB1,SheB2}, while for
$n=2\delta-1$, it reduces to the modified Friedmann equation in
Tsallis cosmology \cite{SheT1}. The second modified Friedmann
equation can be easily derived by combining the first modified
Friedmann equation (\ref{Fried5}) with the continuity equation
(\ref{Con}).

Let us emphasize the distinction between the approach outlined
here and those discussed in Refs. \cite{Luci1,Luci2}. The authors
of \cite{Luci1,Luci2} have altered the total energy density within
the Friedmann equations. Their derived Friedmann equations
resemble the standard form but include an additional dark energy
component that accounts for the effects of the corrected
mass-to-horizon entropy. In contrast, our approach modifies the
entropy in a way that impacts the geometric (gravitational) aspect
of the cosmological field equations, while the energy content of
the universe remains unchanged. From a physical perspective, this
approach is justified, as entropy fundamentally depends on the
geometry of spacetime (the gravitational component of the action).
Consequently, any alteration to the entropy should directly
influence the gravitational side of the dynamic field equations.
\subsection{Generalized Second law of thermodynamics\label{GSL}}
Next, we will examine the validity of the generalized second law
of thermodynamics when the entropy associated with the horizon is
defined by  mass-to-horizon entropy (\ref{Ent1}). This
investigation will take place within the framework of an
accelerating universe, where the generalized second law of
thermodynamics has been previously explored
\cite{wang1,wang2,SheyGSL}.

Combining Eq. (\ref{Fried2}) with continuity equation yields
\begin{equation} \label{dotr1}
\dot{\tilde {r}}_{A}=\frac{4\pi GH}{n\gamma}\tilde{r}_{A}^{4-n}
(\rho+p).
\end{equation}
When the dominant energy condition holds, $\rho+p\geq0$, we have
$\dot{\tilde{r}}_{A}\geq0$. Let us now calculate $T_{h}
\dot{S_{h}}$. It is easy to show that
\begin{eqnarray}\label{TSh1}
T_{h} \dot{S_{h}}&=&4\pi H \tilde{r}_{A}^3  (\rho+p)
\left(1-\frac{\dot {\tilde r}_A}{2H\tilde r_A}\right).
\end{eqnarray}
The violation of the dominant energy condition, represented by the
inequality  $\rho+p<0$, implies that the condition
$\dot{S_{h}}\geq0$ is no longer valid. In this scenario, it
becomes necessary to consider the time evolution of the total
entropy, which includes both the entropy associated with the
horizon, $S_h$ and the entropy of the matter field within the
universe, denoted as $S_m$. Thus, the total entropy can be
expressed as $S=S_h+S_m$.

The Gibbs equation implies \cite{Pavon2}
\begin{equation}\label{Gib2}
T_m dS_{m}=d(\rho V)+pdV=V d\rho+(\rho+p)dV.
\end{equation}
The temperature and entropy of the matter fields within the
universe are represented by $T_{m}$ and $S_m$, respectively. We
suggest that the boundary of the universe is in thermal
equilibrium with the matter field inside it, which means that the
temperatures of both components are equal, i.e., $T_m\simeq T_h$
\cite{Pavon2}. If we relax the local equilibrium hypothesis, it
would lead to an observable energy flow between the horizon and
the bulk fluid, a situation that is not physically acceptable.
According to the Gibbs equation (\ref{Gib2}), one can express this
relationship as follows.
\begin{equation}\label{TSm2}
T_{h} \dot{S_{m}} =4\pi {\tilde{r}_{A}^2}\dot {\tilde
r}_A(\rho+p)-4\pi {\tilde{r}_{A}^3}H(\rho+p).
\end{equation}
Next, we consider the time evolution of the total entropy $S_h +
S_m$. Combining Eqs. (\ref{TSh1}) and (\ref{TSm2}),  we arrive at
\begin{equation}\label{GSL2}
T_{h}( \dot{S_{h}}+\dot{S_{m}})=2\pi{\tilde r_A}^{2}(\rho+p)\dot
{\tilde r}_A.
\end{equation}
Combining $\dot {\tilde r}_A$ from Eq. (\ref{dotr1}) with
(\ref{GSL2}), we finally arrive at
\begin{equation}\label{GSL3}
T_{h}( \dot{S_{h}}+\dot{S_{m}})=\frac{8 \pi^2 G H}{n\gamma}
{\tilde r_A}^{6-n}(\rho+p)^2\geq0.
\end{equation}
In conclusion, when the horizon entropy takes the form of the
generalized mass-to-horizon entropy as described in equation
(\ref{Ent1}), the generalized second law of thermodynamics is
satisfied for a universe that is bounded by the apparent horizon.
\section{Emergence of the cosmic space through mass-to-horizon entropy \label{Emer}}
In this section, we apply the gravity emergence scenario proposed
by Padmanabhan \cite{PadEm} to derive corrections to the Friedmann
equation based on the generalized mass-to-horizon entropy
expression presented in Eq. (\ref{Ent1}).  According to
Padmanabhan, in a pure de Sitter universe characterized by the
Hubble constant $H$, the holographic principle can be expressed as
$N_{\rm sur}=N_{\rm bulk}$, where $N_{\rm sur}$, and $N_{\rm
bulk}$, represent the degrees of freedom on the boundary and in
the bulk, respectively. For our actual universe, which is
asymptotically de Sitter as supported by numerous astronomical
observations, Padmanabhan proposed that the increase in cosmic
volume  $dV$ during an infinitesimal interval $dt$ of cosmic time
is given by \cite{PadEm}
\begin{equation}
\frac{dV}{dt}\propto
\left(N_{\mathrm{sur}}-N_{\mathrm{bulk}}\right). \label{dV1}
\end{equation}
For a flat universe, Padmanabhan assumed the temperature and
volume as $T=H/2\pi$ and $V=4\pi/3H^3$. The reason for this
assumption comes from the fact that in this case one may consider
our universe as an asymptotically de Sitter space. Mathematically,
Padmanabhan proposed \cite{PadEm}
\begin{equation} \label{dV}
\frac{dV}{dt}=G(N_{\mathrm{sur}}-N_{\mathrm{bulk}}).
\end{equation}
Following Padmanabhan, the notion was also extended to a nonflat
universe where it was shown that the Friedmann equations in
Einstein, Gauss-Bonnet and more general Lovelock gravity with any
spatial curvature can be derived by applying the emergence
scenario to the apparent horizon \cite{Sheyem}. It was argued that
in this case one should replace the Hubble radius ($H^{-1}$) with
the apparent horizon radius $ \tilde
{r}_{A}=1/{\sqrt{H^2+k/a^2}}$, which is a generalization of Hubble
radius for $k\neq0$. The generalization of Eq. (\ref{dV}), for a
nonflat universe was proposed as \cite{Sheyem}
\begin{equation}\label{dV1}
\frac{dV}{dt}=G\frac{\tilde {r}_{A}}{H^{-1}}
\left(N_{\mathrm{sur}}-N_{\mathrm{bulk}}\right).
\end{equation}
The temperature associated with the apparent horizon is assumed to
be \cite{CaiLi}
\begin{equation}\label{T2}
T=\frac{1}{2\pi  \tilde {r}_{A}}.
 \end{equation}
The choice to use this temperature expression instead of relation
(\ref{T}) is based on our intention to analyze an equilibrium
system \cite{CaiLi}. Therefore, we propose that within an
infinitesimal time interval  $dt$, the condition $\dot {R}\ll 2H
\tilde {r}_{A}$, holds. This implies that the radius of the
apparent horizon remains effectively constant during this brief
period, akin to the conditions found in a de Sitter universe
\cite{CaiEm}. Padmanabhan's proposal indeed connects the change in
volume $dV$ during this infinitesimal interval $dt$ of cosmic time
to the degrees of freedom present. Consequently, it is justifiable
to disregard the dynamic terms in the Hayward surface gravity,
allowing us to approximate it as $\kappa \simeq {1}/{\tilde
{r}_{A}}$. This simplification leads to the well-known expression
for the horizon temperature. Furthermore, since our universe is
considered to be asymptotically de Sitter, we should adopt the
temperature as expressed in Eq. (\ref{T2}). This assumption is
crucial for deriving the correct form of the Friedmann equations
within Padmanabhan's framework. Additionally, in the context of
Padmanabhan's emergent gravity paradigm, the relation for volume
change assumes that the system is in a state of near thermal
equilibrium at each infinitesimal time step. In this framework,
treating the horizon radius as effectively constant during this
short interval is both physically meaningful and aligns with the
principles of horizon thermodynamics in slowly varying spacetimes.
It is worth noting that in section \ref{First}, one could also
consider the temperature associated with the apparent horizon in
the form of Eq. (\ref{T2}); however, in that case, the first law
of thermodynamics should be applied as $dQ=TdS$, where $dQ=-dE$
represents the energy flux crossing the horizon, and the volume
term should be excluded from this first law \cite{Cai2}.

Using the entropy expression (\ref{Ent1}), we define the number of
degrees of freedom on the surface as
\begin{eqnarray} \label{Nsur2}
N_{\mathrm{sur}}=\frac{8\pi n \gamma}{G(3-n)} \tilde
{r}_{A}^{n+1}.
\end{eqnarray}
With this definition, the surface degrees of freedom are still
proportional to the generalized entropy $S_h$, but the
proportionality constant is chosen so that the resulting effective
gravitational constant matches the one derived from the first law
of thermodynamics. For $n=\gamma=1$, Eq. (\ref{Nsur2}) reduces to
the standard relation $N_{\rm sur}=4S_{h}$.

We also modify the Padmanabhan's proposal as
\begin{equation}\label{dV2}
\frac{d\tilde{V}_n}{dt}=G\frac{\tilde {r}_{A}}{H^{-1}}
\left(N_{\mathrm{sur}}-N_{\mathrm{bulk}}\right),
\end{equation}
where the effective volume is defined as $\tilde {V}_n=\alpha
\tilde {r}_{A} ^{n+2} $. Here $\alpha$ is a constant which for
latter convenience, we choose it as
\begin{equation}\label{alpha}
\alpha=\frac{4\pi n \gamma }{n+2}.
\end{equation}
Clearly for $n=\gamma=1$, we have $\alpha=4\pi/3$ and $\tilde{V}_n
\rightarrow V=4\pi \tilde {r}_{A}^3/3$. The motivation for
choosing the effective volume $\tilde {V}_n$ instead of the usual
volume, comes from the fact that for the generalized
mass-to-horizon entropy $S_h\sim \tilde {A}_{n}\sim \tilde
{r}_{A}^{n+1} $. Thus, the generalized volume corresponding to the
generalized area $\tilde {A}_{n}$ is expected to be $\tilde {V}_n
\sim \tilde {r}_{A} ^{n+2}$.

We take the total energy contained within the apparent horizon as
the Komar energy,
\begin{equation}
E_{\mathrm{Komar}}=|(\rho +3p)|V.  \label{Komar}
\end{equation}
The number of degrees of freedom of the matter field in the bulk
is determined using the equipartition law of energy ($k_B=1$),
\begin{equation}
N_{\mathrm{bulk}}=\frac{2|E_{\mathrm{Komar}}|}{T}.  \label{Nbulk}
\end{equation}
Combining this relation with Eq. (\ref{Komar}) and assuming, in an
expanding universe, $\rho+3p<0$, we find
\begin{equation}
N_{\rm bulk}=-\frac{16 \pi^2}{3}  \tilde {r}_{A}^4 (\rho+3p).
\label{Nbulk}
\end{equation}
Substituting relations (\ref{Nsur2}) and (\ref{Nbulk}) in
assumption (\ref{dV2}), after simplifying, we arrive at
\begin{eqnarray}
\frac{\alpha (n+2)}{4\pi H}\tilde {r}_{A}^{n-4}\dot{\tilde
{r}}_{A}-\frac{2n\gamma}{3-n}\tilde {r}_{A}^{n-3}=\frac{4\pi G
}{3}(\rho+3p). \label{Frgb11}
\end{eqnarray}
If we multiply both side of Eq. (\ref{Frgb11}) by factor
$2\dot{a}a$, after some algebra and using continuity equation
(\ref{Cons}), we reach
\begin{equation}\label{Frgbd2}
\left(\frac{2 n\gamma}{3-n}\right)\frac{d}{dt} \left(a^2 \tilde
{r}_{A}^{n-3} \right)=\frac{8 \pi G }{3} \frac{d}{dt}(\rho a^2).
\end{equation}
Integrating yields
\begin{equation}\label{Frgb3}
\left(H^2+\frac{k}{a^2}\right)^{(3-n)/2} = \frac{8\pi G_{\rm
eff}}{3} (\rho+\rho_{\Lambda}),
\end{equation}
where in the last step, we have used relation (\ref{radius}).
Here, $G_{\rm eff}$ is the effective gravitational constant given
by Eq. (\ref{Gef}).

To sum up, we have derived the modified Friedmann equation
inspired by the generalized mass-to-horizon entropy using the
framework of emergent gravity proposed in \cite{PadEm} and
developed in \cite{Sheyem}. It is straightforward to verify that
the results obtained here align with those from the previous
section; they are identical. Consequently, our findings further
reinforce the validity of Padmanabhan's perspective on emergent
gravity.
\section{Cosmological implications \label{cosm}}
Based on the modified Friedmann equation presented in the previous
sections, we can study the cosmological implications, focusing
specifically on the matter-dominated era in a flat universe.
Therefore, we neglect the contribution from radiation and
cosmological constant.

For a flat, matter-dominated universe ($\rho_{\Lambda}\approx0,
k=0$), the modified Friedmann equation simplifies to
\begin{equation}\label{Fr1}
H^{3-n} = \frac{8\pi G_{\rm eff}}{3} \rho.
\end{equation}
From the continuity equation for the pressureless matter
($p_{m}=0$), we have $\rho_{m}=\rho_{m,0} a^{-3}$, where
$\rho_{m,0}$ is the present matter density, we can substitute to
get
\begin{equation}\label{Fr2}
H^{3-n} = \left(\frac{8\pi G_{\rm eff}}{3} \rho_{m,0}\right)
a^{-3},
\end{equation}
which can be rewritten as
\begin{equation}\label{Fr3}
H = \left(\frac{8\pi G_{\rm eff}}{3} \rho_{m,0}\right)^{1/(3-n)}
a^{\frac{-3}{3-n}}.
\end{equation}
Integrating gives the scale factor as a function of time,
\begin{equation}\label{a1}
a(t)=C_2 t^{(3-n)/3},
\end{equation}
where $C_2=\left(\frac{3 C_1}{3-n}\right)^{(3-n)/3}$, and
$C_1=\left(\frac{8\pi G_{\rm eff}}{3}\right)^{1/(3-n)}$. When
$n=1$, we recover the standard result: $H \propto a^{-3/2}$, or
$a\propto t^{2/3}$. For $n\neq1$, however, the expansion rate
differs from the standard model. In this case we have $a\propto
H^{(n-3)/3}$. Let us study the cases $n>1$ and $n<1$ separately.
(i) For $n <1$, the exponent  $-3/(3- n)>-3/2$. This means $H$
decays slower with expansion than in standard cosmology. The
universe expands faster for a given scale factor. (ii) For $n
>1$, the exponent  $-3/(3-n)<-3/2 $. The exponent $-3/(3-n
)<-3/2 $. This means $H$ decays faster with expansion. The
universe expands slower for a given scale factor.

This model also provides a geometric origin for the accelerated
expansion without invoking any kind of dark energy, and even
without a cosmological constant ($\rho_{\Lambda}=0$). The modified
expansion law for $n\neq1$ introduces terms that do not scale like
standard matter. To get acceleration, we need $\ddot{a}(t)>0$.
From (\ref{a1}), we find
\begin{equation}\label{addt}
\ddot{a}(t)=\frac{n(n-3)}{9}C_2 t^{(-n-3)/3}.
\end{equation}
Thus, for either $n<0$ or $n>3$, we have $\ddot{a}(t)>0$ provided
$C_2>0$. However, since $n>3$, can lead to $C_2<0$, thus the
condition  for an accelerated expansion implies $n<0$. Besides,
the scale factor $a(t)$ should be an increasing function of time
$t$, thus relation (\ref{a1}) implies that $n<3$. Therefore, in
the absence of cosmological constant, this model can explain an
accelerated universe for $n<0$.

In conclusion, the generalized mass-to-horizon entropy model has
rich and testable cosmological implications. During the
matter-dominated era, it predicts a non-standard expansion history
and a modified effective gravitational strength. These deviations
leave imprints on observable phenomena such as the evolution of
the Hubble parameter, the age of the universe, and the large-scale
structure of the cosmos, providing a direct means to constrain the
parameters  $n$ and $\gamma$ with observational data.
\section{Closing remarks \label{Con}}
In this work, we have successfully constructed a unified
thermodynamic framework for the emergence of cosmic space,
anchored in a generalized mass-to-horizon entropy relation
$M=\gamma \frac{c^2}{G}L^n$. We demonstrated the robustness of
this approach by deriving the modified Friedmann equations through
two independent yet convergent thermodynamic routes.

First, by applying the first law of thermodynamics, $dE = T_h dS_h
+ WdV$, to the apparent horizon endowed with the generalized
entropy $S_{h}=\frac{2\pi n \gamma}{G(n+1)}\tilde{r}_{A}^{n+1}$,
we obtained a modified Friedmann equation whose form depends on
the entropic exponents $n$ and $\gamma$. The obtained Friedmann
equation restores the special cases such as modified Friedmann
equations in Tsallis and Barrow cosmology by suitable choice of
the parameters  $n$ and $\gamma$. Subsequently, we embedded the
same entropy relation into Padmanabhan's emergent gravity
paradigm. By redefining the surface degrees of freedom
$N_{\mathrm{sur}}$  and the effective volume $\tilde{V}_n$,
consistent with our entropy ansatz, we independently recovered the
same modified cosmological dynamics.

The remarkable consistency between the results from the
equilibrium thermodynamics perspective and the dynamic emergence
scenario provides strong, cross-validated support for the
mass-to-horizon entropy as a fundamental bridge. It connects the
information-theoretic microstructure of spacetime, encoded in
horizon entropy, to the effective cosmological description
described by the Friedmann equations. Furthermore, we have shown
that the generalized second law of thermodynamics is rigorously
satisfied for a universe bounded by the apparent horizon within
this framework. We explored the evolution of the scale factor in a
flat matter-dominated universe and in the absence of a
cosmological constant. Remarkably, we disclosed that our model
provides a geometric origin for the accelerated expansion without
invoking any kind of dark energy provided $n$ is chosen suitably.

This work not only generalizes previous entropic cosmology models
but also opens a new pathway to understanding cosmic acceleration
and other cosmological phenomena as manifestations of the
non-standard, holographic thermodynamics of spacetime. The
parameters $n$ and $\gamma$, which can be constrained by
observational data, offer a tangible link between
quantum-gravitational insights into entropy and the large-scale
evolution of our universe.
\acknowledgments{I am grateful to Shiraz university Research
Council. I also acknowledge the respected referee for valuable
comments which helped me improve my paper significantly. This work
is based upon research funded by Iran National Science Foundation
(INSF) under project No. 4022705.}



\begin{thebibliography}{99}

\bibitem{Haw1} J. M. Bardeen, B. Carter, and S. W. Hawking, \textit{The Four Laws of
Black Hole Mechanics}, Commun. Math. Phys. {\bf31}, 161 (1973).

\bibitem{Haw2} S. W. Hawking, "Particle Creation by Black Holes," Commun. Math.
Phys. {\bf43}, 199 (1975).

\bibitem{Jac} T. Jacobson, \textit{Thermodynamics of Spacetime: The Einstein Equation of State},
Phys. Rev. Lett. {\bf75}, 1260 (1995), [arXiv:gr-qc/9504004].


\bibitem{Pad1} T. Padmanabhan, \textit{Gravity and the Thermodynamics of Horizons},
Phys. Rept. {\bf 406}, 49 (2005), [arXiv:gr-qc/0311036].

\bibitem{Pad2}  T. Padmanabhan, \textit{Thermodynamical Aspects of Gravity: New insights,}
Rept. Prog. Phys. {\bf73}, 046901 (2010), [arXiv:0911.5004].

\bibitem{Pad3}  A. Paranjape, S. Sarkar, T. Padmanabhan, \textit{Thermodynamic route to Field equations in Lanczos-Lovelock
Gravity}, Phys. Rev. D {\bf74}, 104015 (2006),
[arXiv:hep-th/0607240].

\bibitem{CaiKim} R. G. Cai and S. P. Kim,
\textit{First Law of Thermodynamics and Friedmann Equations of
Friedmann-Robertson-Walker Universe}, JHEP {\bf0502}, 050 (2005),
[arXiv:hep-th/0501055].

\bibitem{Cai2} M.~Akbar and R.~G.~Cai, \textit{Thermodynamic behavior of the Friedmann equation at the apparent horizon of the FRW universe},
Phys. Rev. D {\bf 75}, 084003 (2007), [arXiv:hep-th/0609128].


\bibitem{Shey1} A. Sheykhi, B. Wang and R. G. Cai, \textit{Thermodynamical Properties of Apparent Horizon in Warped DGP Braneworld},
 Nucl. Phys. B {\bf779}, 1  (2007), [arXiv:hep-th/0701198].

 \bibitem{Shey2} A. Sheykhi, B. Wang and R. G. Cai, \textit{Deep connection between thermodynamics and gravity in Gauss-Bonnet braneworlds},
  Phys. Rev. D {\bf76}, 023515 (2007), [arXiv:hep-th/0701261].


\bibitem{PadEm}  T. Padmanabhan, \textit{Emergence and Expansion of Cosmic Space as due to the Quest for Holographic Equipartition},
 [arXiv:1206.4916].

\bibitem{CaiEm}  R. G. Cai, \textit{Emergence of Space and Spacetime Dynamics of Friedmann-Robertson-Walker Universe},
JHEP {\bf11}, 016  (2012), [arXiv:1207.0622].


\bibitem{Yang1}  K. Yang, Y. X. Liu and Y. Q. Wang, \textit{Emergence of Cosmic Space and the Generalized Holographic Equipartition},
 Phys. Rev. D {\bf86}, 104013 (2012), [arXiv:1207.3515].

\bibitem{Sheyem}  A. Sheykhi, \textit{Friedmann equations from emergence of cosmic
space}, Phys. Rev. D {\bf87}, 061501(R) (2013), [arXiv:1304.3054].

\bibitem{Cai3} R.-G. Cai, L.-M. Cao, and Y.-P. Hu,\textit{Corrected
Entropy-Area Relation and Modified Friedmann Equations}, JHEP
{\bf08}, 090 (2008).

\bibitem{sheyECFE} A. Sheykhi, \textit{Entropic corrections to Friedmann equations}, Phys. Rev. D {\bf81}, 104011 (2010), [arXiv:1004.0627].

\bibitem{SheyLog} A. Sheykhi, \textit{Thermodynamics of apparent horizon and modified Friedmann equations},
 Eur. Phys. J. C {\bf69},  265 (2010), [arXiv:1012.0383].

\bibitem{SheyPL} A. Sheykhi and S. H. Hendi, \textit{Power-law entropic corrections to Newton law and Friedmann equations},
 Phys.  Rev.  D {\bf84}, 044023 (2011), [arXiv:1011.0676].


\bibitem{Goh1} H. Gohar and V. Salzano, \textit{On the foundations of
entropic cosmologies: inconsistencies, possible solutions and dead
end signs}, Phys. Lett. B {\bf855}, 138781 (2024),
[arXiv:2307.01768].

\bibitem{Goh2} H. Gohar and V. Salzano, \textit{A generalized
mass-to-horizon relation: a new global approach to entropic
cosmologies and its connection to $\Lambda CDM$}, Phys. Rev. D
{\bf109}, 084075 (2024), [arXiv:2307.06239].

\bibitem{Bas1} S. Basilakos, D. Polarski, and J. Sola,
\textit{Generalizing the running vacuum energy model and comparing
with the entropic-force models}, Phys. Rev. D {\bf86}, 043010
(2012), [arXiv:1204.4806].

\bibitem{Bas2} S. Basilakos and J. Sola, \textit{Entropic-force dark
energy reconsidered,} Phys. Rev. D {\bf90}, 023008 (2014),
[arXiv:1402.6594].

\bibitem{Tsa} C. Tsallis and L. J. L. Cirto, \textit{Black hole
thermodynamical entropy,} Eur. Phys. J. C {\bf73} (2013),
[arXiv:1202.2154].

\bibitem {Bar} J. D. Barrow, \textit{The Area of a Rough Black Hole},
Phys. Lett. B {\bf808}, 135643 (2020), [arXiv:2004.09444].


\bibitem{Bas3} S. Basilakos, A. Lymperis, M. Petronikolou, and E. N.
Saridakis, \textit{Modified cosmology though spacetime
thermodynamics and generalized mass-to-horizon entropy},
[arXiv:2503.24355].

\bibitem{Luci1} G. G. Luciano, \textit{Modified cosmology through generalized
mass-to-horizon entropy, implications for structure growth and
primordial gravitational waves}, [arXiv:2510.00673].

\bibitem{Luci2} G.G Luciano, A. Paliathanasis, \textit{Modified cosmology through generalized mass-to-horizon entropy: observational
constraints from DESI DR2 BAO data}, [arXiv:2508.13260].

\bibitem{Hooft} G.~'t~Hooft, \textit{Dimensional Reduction in Quantum Gravity}, Conf. Proc. C {\bf930308}, 284 (1993),
[arXiv:gr-qc/9310026].
 \bibitem{Sus} L. Susskind, \textit{The World as a Hologram}, J. Math. Phys. {\bf36}, 6377 (1995),
[arXiv:hep-th/9409089].

\bibitem{Noj} S. Nojiri, S. D. Odintsov, and V. Faraoni, \textit{Area-law versus Renyi and Tsallis black hole entropies}, Phys. Rev. D {\bf104},
084030 (2021), [arXiv:2109.05315].

\bibitem{Cimd1} I. Cimidiker, M. P. Da browski, and H. Gohar, \textit{Equilibrium Temperature for Black Holes with Nonextensive Entropy}, Eur. Phys. J. C
{\bf83}, 169 (2023), [arXiv:2208.04473].

\bibitem{Cimd2} I. Cimidiker, M. P. Da browski, and H. Gohar, \textit{Generalized Uncertainty Principle Impact on Nonextensive Black Hole Thermodynamics}, Class. Quant.
Grav. {\bf40}, 145001 (2023), [arXiv:2301.00609].

\bibitem{Ver}  E. Verlinde, \textit{On the Origin of Gravity and the Laws of Newton}, JHEP {\bf1104}, 029 (2011),
[arXiv:1001.0785].


\bibitem{Yang} Y. Gong and A. Wang, \textit{Friedmann Equations and Thermodynamics of Apparent Horizons,} Phys. Rev. Lett. {\bf99}, 211301 (2007),
[arXiv:0704.0793].

\bibitem{Zam} D. J. Zamora and C. Tsallis, \textit{Thermodynamically consistent entropic late-time cosmological acceleration}, Eur. Phys. J. C {\bf82}, 689 (2022),
[arXiv:2201.03385].

\bibitem{Hay1} S.~A.~Hayward, \textit{Unified
first law of black-hole dynamics and relativistic thermodynamics},
Class.\ Quant.\ Grav.\ {\bf 15}, 3147 (1998)
[arXiv:gr-qc/9710089].
\bibitem{Hay2} S.~A.~Hayward, S.~Mukohyama and M.~C.~Ashworth, \textit{Dynamic black-hole entropy}, Phys.\ Lett.\ A {\bf 256}, 347 (1999) [arXiv:gr-qc/9810006].
\bibitem{Bak} D. Bak and S. J. Rey, \textit{Cosmic Holography,} Class. Quant. Grav. {\bf17}, L83 (2000), [arXiv:hep-th/9902173].


\bibitem{SheB1} A. Sheykhi, \textit{Barrow entropy corrections to Friedmann
equations}, Phys. Rev. D {\bf103}, 123503 (2021),
[arXiv:2102.06550].

\bibitem{SheB2} A. Sheykhi, \textit{Modified cosmology through Barrow
entropy}, Phys.  Rev.  D {\bf107}, 023505 (2023),
[arXiv:2210.12525].


{\bibitem{SheT1} A. Sheykhi, \textit{Modified Friedmann Equations
from Tsallis Entropy}, Phys. Lett. B {\bf785}, 118 (2018),
[arXiv:1806.03996].}

\bibitem{wang1} B. Wang, Y. Gong, E. Abdalla, \textit{Thermodynamics of an accelerated expanding universe},
 Phys. Rev. D {\bf74}, 083520 (2006), [arXiv:gr-qc/0511051].

\bibitem{wang2} J. Zhou, B. Wang, Y. Gong, E. Abdalla, \textit{The generalized second law of thermodynamics in the accelerating universe},
Phys. Lett. B {\bf652}, 86 (2007), [arXiv:0705.1264].

\bibitem{SheyGSL} A. Sheykhi, B. Wang, \textit{Generalized second law of thermodynamics in
GB braneworld,} Phys. Lett. B {\bf678}, 434 (2009),
[arXiv:0811.4478].


\bibitem{Pavon2} G. Izquierdo and D. Pavon, \textit{Dark energy and the generalized second law},
 Phys. Lett. B {\bf633}, 420 (2006), [arXiv:astro-ph/0505601].
\bibitem{CaiLi} R. G. Cai, L.M. Cao, Y. P. Hu,
\textit{Hawking Radiation of Apparent Horizon in a FRW Universe},
Class. Quant. Grav. {\bf26}, 155018 (2009), [arXiv:0809.1554].

\end{thebibliography}
\end{document}